\documentclass{webofc}
\usepackage[varg]{txfonts}   
\begin{document}
\title{Studying baryon production using two-particle angular correlations}
%
%

\author{\firstname{Ma{\l}gorzata Anna} \lastname{Janik}\inst{1}\fnsep\thanks{\email{majanik@cern.ch}} 
}

\institute{Faculty of Phyics, Warsaw University of Technology, ul. Koszykowa 75, 00-662 Warsaw, Poland
          }

\abstract{%

Latest measurements of $\Delta\eta\Delta\varphi$  correlations of identified particles show differences in particle production between baryons and mesons. The correlation functions for mesons exhibit the expected peak dominated by effects of mini-jet fragmentation and are reproduced well by general purpose Monte Carlo generators. For baryon pairs (where both particles have the same baryon number) a surprising near-side anti-correlation structure is observed instead of a peak, implying that two such particles are rarely produced with similar momentum. These results present a challenge to the contemporary models and there is no definite theoretical explanation of the observation. In this proceedings an overview of the latest baryon correlation measurements yielding startling results are presented.
}
\maketitle
\section{Introduction}
\label{intro}
Two-particle angular correlations are a robust tool that  enable the exploration of the underlying physics phenomena of particle production in collisions of both protons and heavy ions by studying the distributions of angles in  $\Delta\eta\Delta\varphi$ space (where $\Delta\eta$ is the pseudorapidity difference and $\Delta\varphi$ is the azimuthal angle difference between two particles).  These correlations open up the possibility to study a number of mechanisms simultaneously. Many phenomena, including mini-jets, elliptic flow, Bose-Einstein correlations, resonance decays, conservation laws, etc. are sources of correlations. Each one produces a characteristic distribution in $\Delta\eta\Delta\varphi$ space and contributes to the final shape of the correlation function. 
%
As an example, the clear peak in $ (\Delta \eta, \Delta \varphi) = (0,0) $ for pairs of identical mesons originates from jets and Bose-Einstein correlations. Similar peaks observed for correlations of pairs of kaons with opposite charges is a result of resonances (primarily $\phi \rightarrow K^+ K^-$) and (also) jets.

The influence of jets on the shape of the correlation function, and hence, the peak in $ (\Delta \eta, \Delta \varphi) = (0,0) $ is expected for all particle pairs. It is indeed observed for all studied meson pairs, as well as for baryon--antibaryon pairs. However, the experimental results \cite {Adam:2016iwf, Adam:2019sdm} show that this peak is not observed for pairs of identical baryons.
\vspace{2cm}

\section{ALICE results}
\label{sec:data}

The studies performed by the ALICE Collaboration were performed on proton--proton collisions with energy of $\sqrt{s}=7$ TeV, as well as energy of $\sqrt{s}=13$~TeV.
They were carried out for pions, kaons, protons, and $\Lambda$ hyperons, separately for particle--particle and particle--antiparticle pairs. Particles selected for the analysis had pseudorapidity within  $|\eta|<0.8$ and transverse momentum $p_{\rm T} < 2.5$ GeV/$c$.
The particle identification was performed using the TPC and TOF detectors, reaching 99\% purity for pions and protons, and 96\% purity for kaons. $\Lambda$ hyperons were reconstructed using their decay topology $\Lambda(\overline{\Lambda})\rightarrow\mathrm{p}\pi^-(\overline{\mathrm{p}}\pi^+)$, with purity of more than  95\%~\cite{Adam:2016iwf}. 

\subsection{Definition of the correlation function}
Experimental correlation function, as reported by ALICE, can be defined as
\begin{equation}
\label{eq:CorrelationFuntion}
C(\Delta\eta,\Delta\varphi)=\frac{S(\Delta\eta,\Delta\varphi)}{B(\Delta\eta,\Delta\varphi)}.
\end{equation}
$S(\Delta\eta,\Delta\varphi)$ is a correlated pairs distribution, while $B(\Delta\eta,\Delta\varphi)$ is a reference distribution, obtained using the event mixing technique.  Both $S$ as well as $B$ are normalized to the number of pairs in each distribution, therefore reported correlation function is a ratio of probabilities (the chance of observation of two particles in respect to the chance of observation of each of those particles separately). The details of measurement are explained in~\cite{Adam:2016iwf}.

\subsection{Results}
\begin{figure}[ht!]
	\centering
	\includegraphics[width=\textwidth]{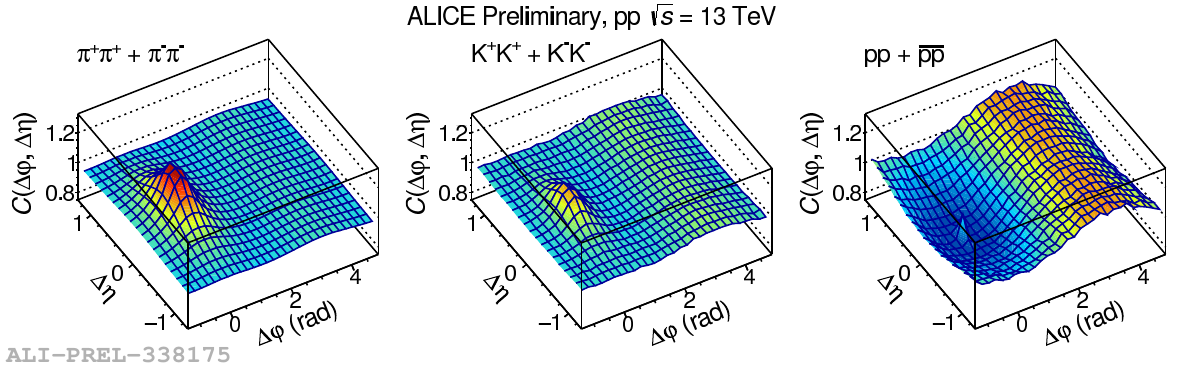}
	\caption{Correlation functions of like-sign pions, kaons, and protons from $\sqrt{s}=13$ TeV  pp collisions measured by ALICE.
	}
	\label{Fig:13TeV_2D}
\end{figure}

Figure \ref{Fig:13TeV_2D} shows correlations for the like-sign pairs of pions, kaons, and protons obtained for $\sqrt{s}$ = 13 TeV pp collisions\footnote{All results for  $\sqrt{s}$ = 7 TeV pp collisions, including correlation functions for pairs of $\Lambda$ hyperons, can be found in \cite{Adam:2016iwf}.} measured by the ALICE experiment.

\begin{figure}[ht]
	\centering
	\includegraphics[width=\textwidth]{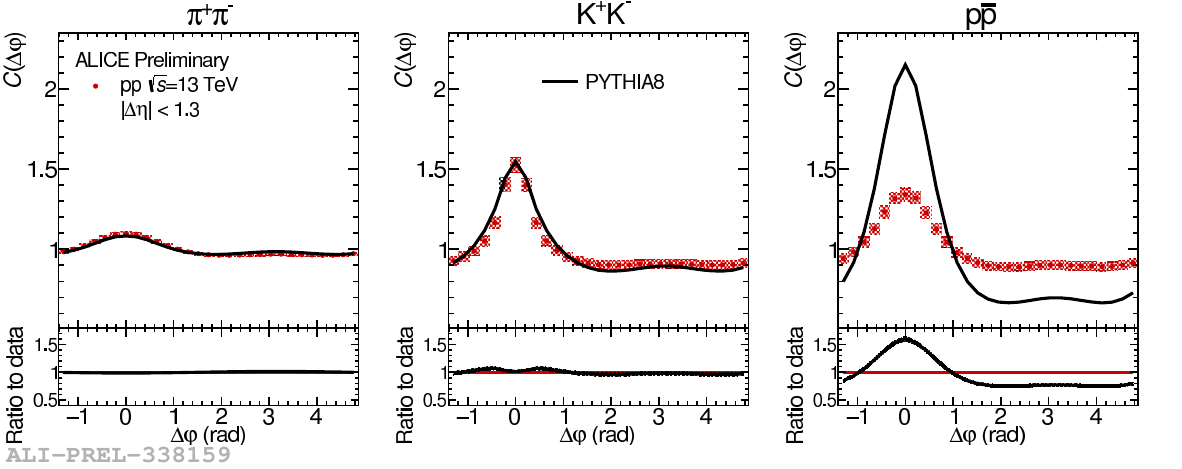}
	\includegraphics[width=\textwidth]{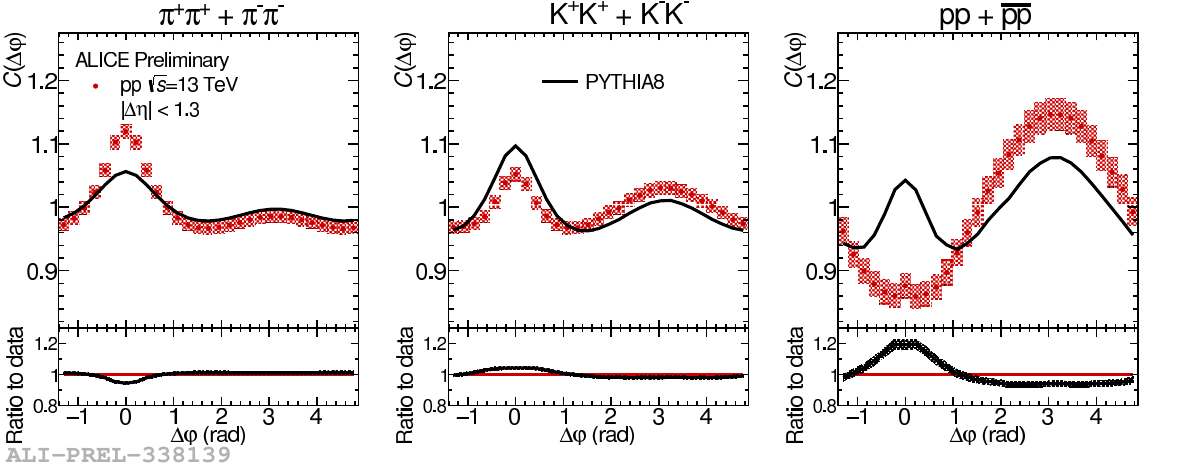}
	\caption{
		$\Delta \eta$ integrated projection of correlation functions for pairs of pions, kaons, and protons obtained from pp collisions at $\sqrt{s}=13$ TeV  from ALICE as well as PYTHIA8 Monte Carlo model. Top panels show unlike-sign particles, while bottom panels -- like-sign particles. Statistical (bars) and systematic (boxes) uncertainties are plotted.}
	\label{Fig:13TeV_1D}
\end{figure}

Results show clear distinct behavior between mesons and baryons: mesons (pions and kaons, left and middle panels) show characteristic near-side peak in ($\Delta\eta,\Delta\varphi) = (0,0)$, 
while for the pairs of identical protons (right panel) we can observe an anti-correlation. Moreover, in Figure \ref{Fig:13TeV_1D}  $\Delta\eta$ integrated projections of the two-dimensional correlation functions are shown for both unlike-sign (top row) and like-sign (bottom row) pairs of pions, kaons and protons, together with PYTHIA8 \cite{Skands:2014pea} model simulations. The majority of Monte Carlo generators\footnote{Comparisons with other models can be found in \cite{Adam:2016iwf}.} do not describe the anti-correlation shape visible fo baryon--baryon pairs, as well as do not quantitatively describe baryon--antibaryon pairs. The most promising are results from the AMPT generator, see \cite{Zhang:2018ucx,Zhang:2019bkf}.

\clearpage
\section{STAR results}
\label{sec:data}
Measurements performed by the STAR experiment were done on $Au-Au$ collisions at $\sqrt{s_{\rm NN}}=7.7$--200~GeV. Analysis was performed for pions, kaons, and protons (separately for different charge combinations) as well as for different centrality ranges.

\subsection{Correlation function definition}
Results of the STAR Collaboration were reported as $R_2$ quantity, which is defined as follows:
\vspace{0.25cm}
\begin{equation}
\label{eq:CorrelationFuntion}
R_2(\Delta y,\Delta\varphi)= \frac{\left \langle n\right \rangle^2}{\left \langle n(n-1)\right \rangle} \frac{\rho_2(\Delta y,\Delta\varphi)}{\rho_1(y_1,\varphi_1)\rho_1(y_2,\varphi_2)} -1,
\end{equation}
where $\Delta y = y_1-y_2$ is the difference between rapidity of two particles,  $n$ stands for multiplicity of the given particle type in the collision, while $\rho_2$ and $\rho_1$ are the multiplicity density distributions of pairs of particles of single particles, respectively, normalized to the number of events.

\subsection{Results}

\begin{figure}[ht!]
	\centering
	\includegraphics[width=\textwidth]{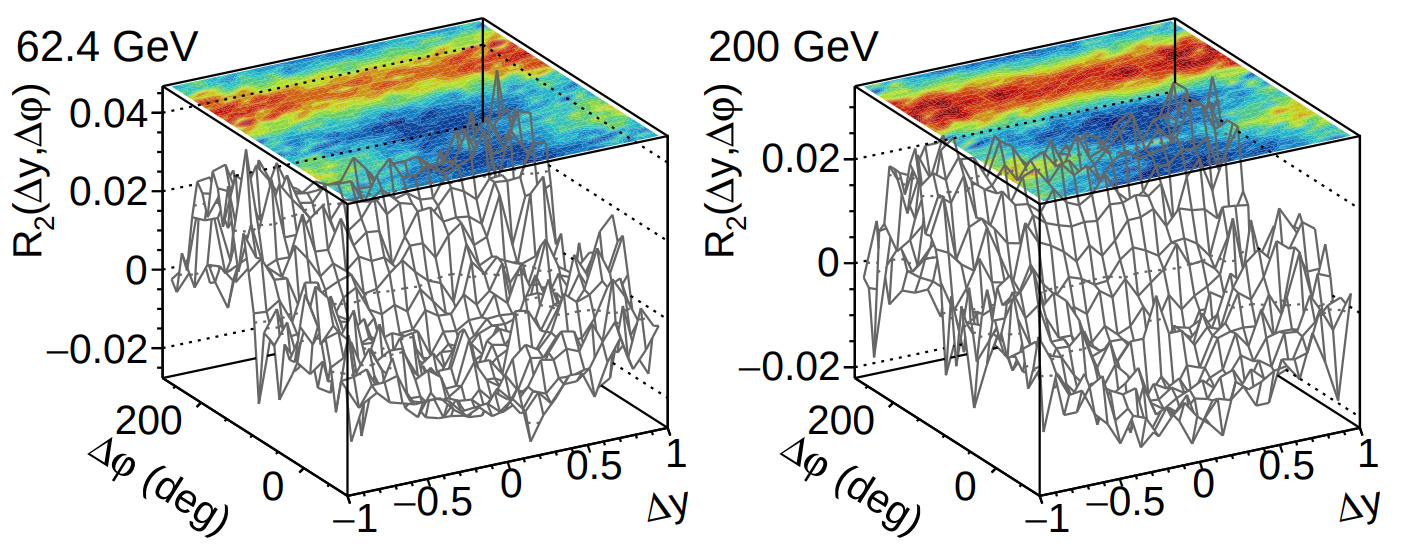}
	\caption{Correlation functions of like-sign protons from  \mbox{Au--Au} collisions at $\sqrt{s_{NN}}=62.4$ and 200 TeV measured by STAR experiment. Figure from  \cite{Adam:2019sdm}.
	}
	\label{Fig:STAR}
\end{figure}

In Figure~\ref{Fig:STAR} distributions for the like-sign pairs of protons are presented for two collision energies: $\sqrt{s_{\rm NN}}= 62.4$ GeV and $\sqrt{s_{\rm NN}}= 200$ GeV (more results can be found in Ref.~\cite{Adam:2019sdm}).
For both distributions previously discussed anti-correlation at  $(\Delta y,\Delta\varphi)=(0,0)$ is clearly visible.
The same distribution shape was observed for all studied energies. 

Similar anti-correlation is visible also for baryon--antibaryon pairs  \cite{Adam:2019sdm}  in heavy-ion collisions, however, of much smaller range; that anti-correlation comes from matter--antimatter annihilation and was observed also in the results of femtoscopy \cite{Acharya:2019ldv} measurements as well as Monte Carlo models ~\cite{Adam:2019sdm}.

\newpage
\section{Discussion}

Different reasons for the observed shape correlation functions were studied. Possible physics effects that could influence baryon--baryon correlation functions were analyzed:

\begin{itemize}
	\item The influence of the Coulomb interaction is negligible: the shape of the correlation functions is very similar for proton--proton and lambda--lambda ($\Lambda$ hyperons are neutral) pairs, regardless of the electric charge of particles \cite{Adam:2016iwf}.
	
	\item Fermi-Dirac quantum statistics effects cannot be the sole reason of the observed anti-correlation. The anti-correlation can be observed for pairs of pp, $\Lambda \Lambda$ (identical particles), and p$\Lambda$ (non-identical particle pairs, that are not influenced by quantum statistics effects) \cite{Adam:2016iwf}.
	
	\item The influence of the momentum of particles was studied; regardless of the momentum the anti-correlation was still observed \cite{Adam:2016iwf}.
	
	\item The dependence on the collision energy was studied as well: for both lower energies \cite{Adam:2019sdm} as well as  LHC energies ($\sqrt{s}=7$ TeV \cite{Adam:2016iwf} and $\sqrt{s}=13$ TeV) anti-correlation is visible.
	
	\item 	The local conservation of baryon number is not the only reason of the anti-correlation: all studied Monte Carlo models included this effect in the calculations, however, they were unable to reproduce the experimental results \cite{Sjostrand:2018xcd}.
\end{itemize}

\section{Summary}
Experimental results, which are not well reproduced by current theoretical models, suggest that the production mechanism of protons -- one of the most common particles in the Universe -- is not well understood. The surprising effect of anti-correlation was observed first for \mbox{$\sqrt{s}=7$ TeV} pp collisions by ALICE, but now we know as well, that the same behavior of like-sign baryons is observed for higher, as well as lower energies, and also for different systems.

\section*{Acknowledgments}
This work was supported by the Polish National Science Centre under decisions no. UMO-2016/22/M/ST2/00176, no. UMO-2017/27/B/ST2/01947 and Polish Ministry of Science and Higher Education.
\vspace{-0.3cm}
%
%
%
\bibliography{biblio}

\end{document}